# Drone Delivery Systems and Energy Management: A Review and Future Trends

**Mohammad Sadra Rajabi**

School of Civil Engineering, Faculty of Engineering, University of Tehran, Tehran, Iran

Msadra.rajabi@ut.ac.ir

**Pedram Beigi**

Department of Civil Engineering, Sharif University of Technology, Tehran, Iran

Beigi.Pedram@gmail.com

**Sina Aghakhani**

Department of Industrial Engineering, Sharif University of Technology, Tehran, Iran

sinaaghakhani1996@gmail.com

## Abstract

Advanced technological breakthroughs and exceptional levels of innovation are enhancing the capabilities and potential of autonomous unmanned aerial vehicles (UAVs or drones), and in so doing attracting the interest of a broader swath of logistic companies, online retailers, and governmental agencies. These technology advancements and their impact on regulatory agencies may soon pave the way for the widespread use of drones for delivery and monitoring purposes. Moreover, increasingly urgent environmental factors that include $CO_2$ emissions reductions and other energy-saving approaches are intensifying to need to reduce vehicular usage and congestion, which could further spur their usage. To optimize these systems, drones often employ a hybrid power supply system architecture to boost endurance and performance. Fuel cells, batteries, solar cells, and supercapacitors are examples of power sources that may be combined in a hybrid power architecture. To enable today's drones (and those of the future) to work efficiently, the appropriate energy management system must be selected based on optimal and accurate modeling techniques This chapter provides a comprehensive review of drone energy-supply management and strategic systems to identify their plusses and minuses, as well as suggests recommendations for future research.

**Keywords**

Unmanned Aerial Vehicles, Energy Consumption of UAVs, Energy Management, Drone Delivery

## 1. Introduction

During recent years, last-mile delivery, as the last phase in the supply and distribution chain, poses a significant challenge due to the growth of emerging delivery systems (Aghakhani *et al.*, 2022). Companies are becoming more innovative in how they transport goods, and as a result of just-in-time management, delivery services are increasing (Beigi, Haque, *et al.*, 2022; Beigi, Khoueiry, *et al.*, 2022). Unmanned aerial vehicles (UAVs) are flying robots that can operate autonomously or telemetrically to perform special missions. Due to ongoing advancements in microprocessors and artificial intelligence (AI), significant improvements in the cost of UAVs and their expanded mobility are now possible, thus capturing the interest of users and researchers in their potential across a range of applications (Adnan *et al.*, 2019; Moeinifard *et al.*, 2022). Notably, many military and civil uses exist for these devices, including delivery services, minesweeping, agriculture-related applications (e.g., pesticide spraying or performing field soil analysis), wireless coverage, and monitoring (Famili *et al.*, 2022a, 2022b, 2022c; Razzaghi and Assadian, 2015). In order to maximize their use of UAVs, several multinational companies are investing heavily in improving UAV performance toward augmenting their utility.

Increasingly, researchers are targeting the energy efficiency of UAVs, which is critical for their performance. Indeed, the development of more judicial energy-management policies that focus on energy conservation has gained considerable attention (Erfani *et al.*, 2021; Erfani and Cui, 2022). Specifically, greater attention is being given to solving significant challenges in drone-delivery systems, such as how to generate and store fuel in smaller fixed-wing UAVs. In fact, a number of current research trends and developmental investigations involving UAVs across different civil applications are addressing their energy-related engineering challenges for enhanced usage (e.g., extending battery life). Other research efforts have focused on classifying and identifying the range of design challenges associated with UAVs, developing path-planning algorithms for fixed-wing UAVs, enhancing the guidance and control of rotorcraft unmanned aircraft, and pursuing quadrotor modeling and management (Gong and Verstraete, 2017a; Kendoul, 2012; Shraim *et al.*, 2018). Based on the current research, this chapter presents a thorough analysis of the energy management systems and power supply architectures of UAVs, with a focus on how ongoing technical developments may impact their utility.

## 2. Drone Energy Suppliers

Despite the application of gas turbine engines in propulsion systems for aircraft, which is characterized by an advantageous power-to-weight ratio and extended operating times, such engines tend to perform better at higher power levels (i.e, above 100 horsepower). Because of their low fuel efficiency and high noise level, they are unsuitable for incorporation in smaller-scale UAVs (Ahmed F. El-Sayed, 2017; Austin, 2010; Hassanalian and Abdelkefi, 2017).

The internal combustion engine (ICE) has long served as the backbone for aircraft propulsion systems. Due to its higher fuel energy and power density, an ICE, in contrast to an electric motor (EM), provides longer flight times and a greater payload range, which are important features when it comes to airborne applications. It must be noted, however, that the multistep process of generating energy reduces the system efficiency of ICEs (Brown *et al.*, 2015; Sharaf and Orhan, 2014). Many key advantages of EMs make them appropriate for UAVs, including their low thermal and acoustic signatures, well-developed electronic controls, the relative ease of adapting them to automatic control, low cost, self-starting features, and improved reliability—the latter being particularly important for reducing the probability of crashed caused by motor failure. Electric propulsion systems are prone to electronic speed controller (ESC) failures caused by overheating, which can result in melting the ESC casing. As recent reports have indicated, duplicating components may alleviate this problem (Brown *et al.*, 2015; Glassock *et al.*, 2009).

Recently, a hybrid UAS prototype that combines thermal and electric engines was proposed that utilizes the advantages of both. Despite simulation results confirming a 13% improvement in endurance, the system is still complex and not environmentally friendly (Glassock *et al.*, 2009). Similarly, other researchers have discussed using EM and ICE in a hybrid powertrain architecture (Bongermino, Mastrorocco, *et al.*, 2017; Bongermino, Tomaselli, *et al.*, 2017; Xie *et al.*, 2019). UAVs currently do not incorporate an ICE since it is not a viable solution for optimizing fuel usage and endurance. This section discusses electrical power sources for UAVs and the potentially intriguing supply strategies for one-source-based UAVs.

### a. Battery-based supplying techniques

#### i. UAVs powered by batteries

A significant proportion of smaller UAVs (e.g., the quadrotors) are battery-powered, highlighting the importance of this portably energy source for UAVs. The use batteries simplifies and improves the flexibility of propulsion systems (Khofiyah *et al.*, 2018). In addition, few would argue that battery-based platforms are fairly cost-effective and provide long flight times, which are important considerations for hobbyists. Nonetheless, the average small battery-powered UAVs may suffer from short endurance as a result of pack weight limitations. When LiPo batteries are incorporated, the UAV can fly for 90 minutes on one charge (Verstraete, Lehmkuehler, *et al.*, 2012). Due to this feature, these smaller UAVs are typically used for commercial purposes. Unlike many other battery types, lithium batteries have a relatively high specific energy, making them ideal for small UAVs. In fact, it is estimated that almost 90% of micro aerial vehicles on the market weigh less than 2 kg, are under 100 cm in length, and are powered by LiPo batteries (Hassanalian and Abdelkefi, 2017).

Several literature reports have examined the parameters that affect the performance of battery-powered UAVs, using mathematical formulas to approximate battery capacity and range. However, neither simulations nor experiments were conducted to evaluate these derived mathematical formulations. Most scientists and engineers would agree that the battery-powered electric vehicle is one of the most challenging vehicles on the road today in terms of operational longevity. As such, intensive research efforts have focused on how to extend a battery's performance and lifetime, thereby allowing electric vehicles to stay on the road for longer. Despite advancements in battery characteristics, the current technology still has a limited range and endurance due to its specific energy (Traub, 2011), making it unable to meet the demands of many UAV applications. In addition, energy density improvements have had an impact on both stability and safety levels. To overcome battery limitations, different approaches have been detailed in literature—with a particular focus on the advantages of fuel cells. Indeed, when examining their performance parameters (e.g., their high specific energy), fuel cells offer a good alternative to batteries and other sources of power for UAVs, such as solar cells and supercapacitors (Mike, 2018).

## ii. Swapping

When UAV batteries become depleted during its flight, swapping is used to recharge them. The process can be autonomous or human-controlled. The term "hot-swapping" refers to replacing the depleted battery with a fully charged one while the UAV is still running; once joining its hotspot, it can operate normally again. Multi-agent systems deliver continuous service through the process of deploying one or more UAVs, and then governing their cooperation such that a hotspot can be seamlessly transferred from one unit to another (Galkin *et al.*, 2019). For a typical swapping operation to be successful, three conditions must be met: (1) the availability of a ground station where UAVs can land for charging or changing batteries; (2) the availability of UAV swarms for persistent applications; (3) the establishment of a reliable management system to facilitate UAV cooperation. Figure 1. illustrates how swapping and hot-swapping methods work.

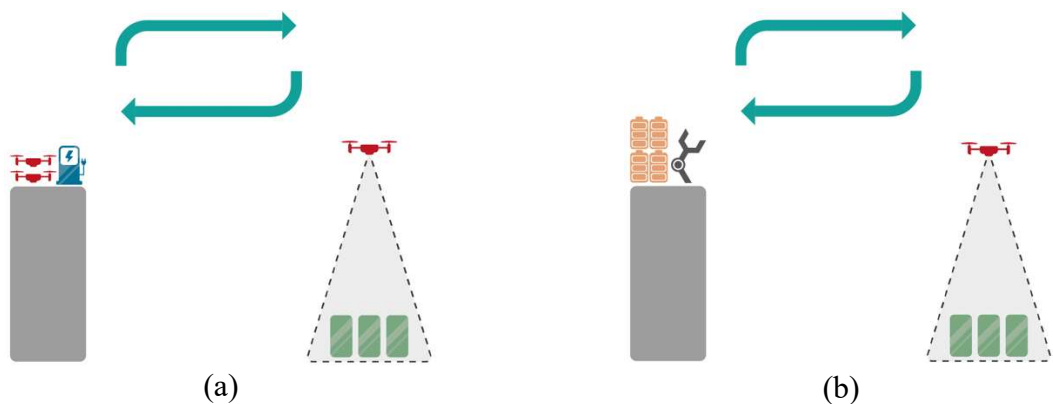

*Figure 1. Swapping (a) and hot-swapping (b) methods*

As part of infrastructure networks, ground stations are established in specified locations such as cities or along trajectories that connect cities ("Neva Aerospace Announces UAV Charging Networks | UAS VISION", n.d.). Researchers have recommended several typical locations for

these infrastructure networks, including power poles, cell towers, streetlights, rooftops, and standalone pylons (Gentry et al., 2016; Saad et al., 2014). In addition to ground electronics, the battery swap station should also incorporate a landing frame, onboard circuit, and contact mechanism (Leonard *et al.*, 2014). Charges can be made via physical paths, magnetic paths, or inductive coupling; additionally, power lines, large batteries, and solar cells can be used as power sources for docking platforms at remote stations.

Recent research has described a multi-rotor aerial prototype for longer-duration missions that utilizes the swapping approach, whereby ongoing monitoring of the battery's state of charge (SoC) is used to maintain the operational status of the airborne platform (Williams and Yakimenko, 2018). This prototype requires that one quadcopter be maintained continuously in the loiter position once the battery level drops below a predefined threshold when the SoC drops below that value. After the second quadcopter takes over, the first one returns to the ground station for recharging and refueling. This system cycles through all UAVs until all batteries have been exhausted or once the mission has been completed. For this study, whether a given number of batteries is sufficient to ensure the durable operation of the system is determined by the charging and discharge times—not by the number of UAVs. However, the larger the system, the better the performance. On the launch platform, changing and charging the batteries was not automated, so the system was still dependent on human intervention. Furthermore, no sensors were used for any kind of special mission; only static loitering was performed during their testing.

Literature reports have compared the economics of different battery refilling/recharging platforms. One study detailed the development of three stations on the basis of an axiomatic design; the authors then analyzed and compared relevant factors such as cost, complexity, and coverage. Their findings indicated that refill stations are best when coverage is poor; otherwise, exchange stations are better (Kemper *et al.*, 2011). Additionally, a swapping system using online algorithms was developed to address energy management, UAV longevity, and accurate landing parameters. Servo-based lifts were utilized in this approach to switch and place the batteries onto a hexagonal mat that was horizontally mounted. However, this swapping mechanism was found to significantly reduces UAV performance (Swieringa *et al.*, 2010).

The use of active infrared imaging for docking platforms has also been reported. This system facilitates a higher-precision landing during the day or at night using a camera and an infrared filter (Liu *et al.*, 2017, 2018). In this approach, image processing is used to guide the UAV landing operation, which was found to reduce the swapping process from 60 seconds to less than 10 seconds. Although published studies do address swapping time and landing precision, cooperation between UAVs is underreported; moreover, improvements in operational times have yet to be fully investigated using long-duration missions (Saha *et al.*, 2011).

In a recent study, Bocewicz et al. investigated cyclically repeated missions, such as aerial deliveries. The authors suggested the used of mobile battery swapping stations (MBSs) that are able to shift to defined swapping locations according to a pre-determined schedule. As a result, UAVs can engage with the most appropriate station(s) for battery replacement and loading/unloading cargo. When considering the challenges involved in MBSs routing, this study suggests a declarative model of routing UAVs and MBSs in an effort to optimize both the number of UAVs and the distance traveled. Despite its potential advantages, this solution would likely only be applicable to a limited range of missions, coupled with the fact that mobile swapping stations would in many cases be impractical. It should also be noted that the study in question focused exclusively on the routing task, but lacked experimental findings for instances when swapping was not implemented (Bocewicz *et al.*, 2019).

### iii. Laser-beam inflight recharging

Based on ground station-related constraints for charging/replacing batteries, the swapping approach has potential as a way to extend the operational parameters of battery-based UAVs, thereby enhancing their flight performance and utility. In this context, a recent report described the use of wireless recharging as another strategy for keeping UAVs flying (Galkin *et al.*, 2019).

Structurally, a ground station is needed to supply power to the laser generator, which transmits a laser beam to the UAV while it is in flight. A receiver embedded in the UAV then converts light into electricity. Accordingly, a UAV equipped with this technology would be able to remain in the air for an indefinite period of time without having to land to refuel (Park *et al.*, 2018). Since an aerial power link area is used to recharge the UAV when it needs power, takeoffs and landings are made safer. On the roofs of high buildings or on mobile stations, laser transmitters can avoid laser-beam obstructions. UAVs and their nearest energy source will be linked by a radiative link to facilitate rapid power transfer. For example, the LaserMotive prototype allows the transfer of hundreds of watts (Achtelik *et al.*, 2011; Ouyang *et al.*, 2018; Shakerian *et al.*, 2022). Further, Ouyang and coworkers demonstrated the technical feasibility of keeping a quadcopter in the air for over 12 hours. Their study was designed to analyze the payload, size, the ability of UAVs to adapt to specific applications, while also detailing their flight-control systems and mechanical design (Ouyang *et al.*, 2018).

When a UAV is to be recharged via a laser beam when it is flying, the apparatus must fly at lower levels; moreover, this approach generally requires the UAV to be operated within a more restricted area so that the laser transmitter can deliver power. As such FAA (Federal Aviation Administration) regulations restrict flying small UAVs above 400 feet ("Small Unmanned Aircraft Systems (UAS) Regulations (Part 107) | Federal Aviation Administration", n.d.). This same regulation also specifies certain weight guidelines: drones used for either commercial or government purposes weighing less than 26 kilograms are permitted. Furthermore, since a single laser transmitter is required for each UAV, the number of UAVs that can be deployed in a given area is limited, which can increase the cost of this approach (Galkin *et al.*, 2019).

### iv. Tethered UAVs

Tethering a UAV to a fixed power source has the advantage of affording it unlimited autonomy. The availability of an ongoing supply of electricity via connection lines from a power supply station negates the need for potentially disruptive repeated recharging on the ground or during flight. Moreover, this approach provides the ground station with a safe and efficient way of transmitting data from the UAV. Typically, a copper cable will serve to transmit power. Copper wires can be reduced eightfold in weight compared to battery payloads and power lines; it must be noted that, however, that power loss with copper wires will reduce efficiency. In contrast, advanced fiber technologies are emerging to provide power to tethered-UAVs; indeed, through the use of fiber optics, thousands of kilowatts of energy can be transferred. Also, by removing the electrical signature, power provided optically will make the signal less detectable. In short, high-altitude deployment of fiber technologies are becoming increasingly beneficial.

Regarding specific operational usages, a tethered UAV is known to be effective for the ongoing monitoring of maritime-related pollution resulting from oil spills, which can be potentially devastating. Woodworth et al. detailed how tethered drones are useful for data gathering across a range of applications (Woodworth and Peverill, 2013). As noted earlier, this approach is inherently hindered by the fact that the UAV's operating area is restricted by the connecting cable that prevents it from flying too far from its base station. However, the UAV can increase its range through the use of a moving vehicle as the main power source. In another recent report, Gu and coworkers described the importance of tethered UAVs in the monitoring of nuclear power plants, for which long-duration surveillance is essential (Gu *et al.*, 2016). For space-restricted assignments such as this, a tethered cable will provide constant power, enabling the UAV to fly for a few days or even months. Several prototypes have been developed and successfully deployed across a number of outdoor settings and situations.

### b. Fuel cell powered UAVs

The use of hydrogen fuel cells to power UAVs is also gaining interest. In comparison to using conventional batteries, a fuel cell significantly improves the in-flight longevity of a UAV (Pan *et al.*, 2019). Lithium-polymer batteries, for instance, have a specific energy of up to 250 Wh/kg (Kim and Kwon, 2012); to compare, the specific energy of a fuel cell system incorporating a compressed hydrogen tank can achieve 1000 Wh/kg (Javadinasr *et al.*, 2021; Verstraete, Lehmkuehler, *et al.*, 2012). Another advantage is while battery recharging can take a long time, the refueling process associated with fuel cells occurs almost instantly.

Literature reports have addressed a number of fuel cell-related factors, notably energy and power densities, efficiency, discharging characteristics, temperature effects, and overall endurance. In particular, researchers have compared and evaluated three battery types: lithium-ion, nickel-cadmium, and nickel-molybdenum (Cai *et al.*, 2010; Donateo *et al.*, 2017; Famili, Stavrou, Wang, Jung-Min and Park, 2021; Famili, Stavrou, Wang, Jung-Min, Park, *et al.*, 2021; Famili and Park, 2020). Despite useful assessments of each type, none of these reports included actual flights that would have provided important data regarding the power sources during flight—such as their capability and performance under various airborne conditions. Therefore, fuel cells should be considered an optimal solution for UAVs with a higher endurance for particular weights (Evangelisti *et al.*, 2017; Javadinasr *et al.*, 2022). In addition to the advantageous volume of the hydrogen, fuel cells feature a lower energy density in comparison to lithium batteries.

Also, Belmonte et al. recent described how UAVs are being used to inspect mobile cranes for defects and weaknesses (Belmonte *et al.*, 2018). Their analysis included a review of UAVs powered by either lithium-ion batteries or proton exchange membrane fuel cells. From an economic perspective, the authors concluded that fuel cells are generally more expensive because they are a niche product.

### i. The drawback of fuel cells efficiency issues

Fuel cells can provide efficiency levels of up to 60%, which is disappointingly lower compared to lithium batteries (over 90%) (Hwang *et al.*, 2013). Operations that entail the use of a fuel cell stack also require auxiliary equipment, which tends to further diminish efficiency as well. Also, the need to generate hydrogen onboard will inevitably increase the complexity of the system (Pan *et al.*, 2019).

### ii. Fuel cell types

Fuel cells employ a variety of technologies. Typically, they are classified in one of two ways: (a) according to their chemical specification (i.e., catalysts and electrolytes), or (b) according to operational parameters such as temperature. Recently, Gong et al. compared the most commonly used fuel cell types in UAVs: the polymer electrolyte membrane fuel cell (PEMFC), the direct methanol fuel cell (DMFC), and the solid oxide fuel cell (SOFC) (Gong and Verstraete, 2017b). As reported in the literature, UAV propulsion systems typically utilize PEMFCs (Pan *et al.*, 2019). (Intelligent Energy is a fuel cell firm that manufactures PEMFCs for UAV applications.) Certain essential characteristics of this fuel cell technology must be considered prior to implementation: their weight, their degree of high-power density, their operating temperature and longevity, and their response time to load variations (Gong and Verstraete, 2017b; Lapeña-Rey *et al.*, 2017).

### iii. Fuel storage

At standard temperature and pressure, hydrogen features a density of only 0.089 kg/m3. As a result, bulky tanks are required to enable the UAV to carry ample fuel for the required mission. This is a critical constraint when considering UAV size and weight. Additionally, pure hydrogen cannot be stored under extremely high pressures or low temperatures for safety reasons. Currently, UAVs use three kinds of hydrogen-storage methods (Gong and Verstraete, 2017b): liquid hydrogen, compressed hydrogen gas, and chemical hydrogen generation. While there are advantages and disadvantages to each of these storage techniques, they are beyond the focus of this paper.

## c. Hybrid power sources

### i. Fuel cell and battery

As an alternative power source for UAVs, fuel cells feature some notable limitations despite their technological advancements and attractive performance parameters. Since fuel and air need to be supplied by pumps, valves, and compressors, a fuel cell has a considerable time constant. There are several reasons why the response is delayed, including the mechanical characteristics of the pump, the flow delay, and thermodynamic characteristics (Ou *et al.*, 2015). Therefore, when the current demand is high, fuel starvation can occur and negatively affect lifetime, reliability, and efficiency (Ou *et al.*, 2018). However, when a fuel cell and battery are combined to create a hybrid power supply system, the distinct advantages of capitalizing on both sources of propulsion outweigh the potential drawbacks (Belmonte *et al.*, 2018; Cooley *et al.*, 2014; Donateo *et al.*, 2017).

Battery power, because it has a higher capacity, is faster and more efficient than fuel cell power. Therefore, this power source is preferred in providing optimal power when the UAV must undertake more energy-intensive maneuvers, notably take-off and climbing. Once energy requirements are less acute, the fuel cell can take over as the primary power source—for example, during cruising and descent. Furthermore, it can maintain the SoC above the recommended threshold by charging the battery.

Through hardware-in-the-loop (HIL) simulations, Verstraete and coworkers assessed the performance of one such hybrid UAV propulsion system, which incorporated a 200 W fuel cell and a battery (Verstraete, Harvey, *et al.*, 2012). Various tests were performed on each source to examine their performance. Hydrogen use and endurance were also measured in this study. Using real flight data recorded during a flight, HIL was applied under different degrees of load fluctuation in another study (Gong *et al.*, 2014). Also, the battery contribution to the hybrid power system has been studied (Gong and Verstraete, 2014). During their experimental investigations they examined the performance of the battery under a variety of conditions encountered during flight mission phases. Additionally, using an array of mission profiles and speeds, Gong and coworkers provided a detailed description of this propulsion system (Gong *et al.*, 2016). One drawback must be noted in connection with these aforementioned studies: Each lack an energy-management strategy for their system due to the fact that power splitting was only conducted passively.

## ii. The use of solar cells as an auxiliary power source

Due to the growing importance of photovoltaic (PV) generation systems, this report will explore in detail the use of PV generation systems deployed in mobile platforms such as UAVs. Using PV arrays on the UAV's wings, the device can fly indefinitely based on solar-generated power and when coupled with the installation of a battery for energy storage during the night (Jaw-Kuen Shiau *et al.*, 2009). In particular, these solar-powered UAVs are ideal for High Altitude Long Endurance (HALE) applications. HALE UAVs are designed to conduct extended missions over longer durations at high altitudes. A solar-powered UAV design method was proposed by Mortons et al. to optimize the efficiency of the airframe. During experimental assays involving the prototype, the authors concluded that the high level of solar energy stored by the UAV outweighed any potential disadvantage of carrying the additional payload of the solar power-generation system (Lotfi *et al.*, 2022; Morton *et al.*, 2015).

Based on Harvey et al.'s findings, PVs may reduce the weight of UAVs by up to 59%, in addition to saving fuel (Harvey *et al.*, 2012). Utilizing solar energy, therefore, is an important strategy for improving UAV endurance, but involves specific design considerations. Importantly, a solar-powered UAV must have large wings if it is to maximize the amount of light energy it receives. Moreover, as reported by Peng and coworkers, a maximum power point tracking (MPPT) algorithm is required, which consists of a simple converter, current and voltage sensors, and a low-cost microcontroller (Peng *et al.*, 2018).

### iii. Supercapacitor as an auxiliary power source

Supercapacitors have received noteworthy attention over the past several years due to the demand for faster energy-storage systems for a growing range of different applications; a supercapacitor can supplant or augment batteries, which are slow to charge and discharge and have a limited lifetime (Aneke and Wang, 2016). Compared to batteries, supercapacitors feature considerably higher power and significantly lower energy densities. Also, this device can function over a broad temperature range, tends to require little maintenance, and is not prohibitively expensive (Ruan *et al.*, 2017). Furthermore, the DC bus voltage fluctuation is greatly reduced as well.

Increasingly, researchers are investigating the incorporation of a supercapacitor in a hybrid power system for an UAV, thereby facilitating an additional range of architectural choices, as well as well as increasing power density and enabling a quicker power response. Fuel cells usually provide the most power in hybrid power supply systems, while other sources act as auxiliary sources. To enhance the lifetime of the UAV, fuel cells are preferred for operating continuously (Li *et al.*, 2018). For each power source to operate optimally, the installation of an electric motor is mandatory.

Based on available literature reports, the researchers have conducted an HIL-based assessment of a hybrid UAV propulsion system that incorporates fuel cells, batteries, and supercapacitors. In order to isolate and understand the role of the supercapacitor, the drone system was compared with a fuel cell/battery system; also studied was the impact of supercapacitor capacity on the efficiency of the fuel cell and supercapacitor. The researchers showed that the supercapacitor performed well—both in terms of load smoothing and dynamic response. Moreover, in some other studies, using the same UAV configuration (Gong *et al.*, 2018) involved actual flight testing with a deployed UAV prototype, with resulting data demonstrating that the supercapacitor can provide optimum power and resolve power fluctuations during dynamic flights involving changing load conditions. Despite the DC bus voltage stabilization also being shown, there was a lack of studies that addressed EMS strategy.

## 3. Energy Management Strategies

The most effective architecture for powering the propulsion system of UAVs is hybridization, which can balance the pros and cons of differing power sources, as well as address their distinct limitations. Consequently, power needs to be split optimally between sources to achieve efficient energy usage and to extend the lives of power sources toward the goal of achieving high performance. To accomplish this objective, power management systems must be implemented in order to split power among the available sources, while at the same time taking into consideration factors such as efficiency, speed, fuel consumption, and required power. This approach involves active power management, whereby the energy management unit controls the power outputs through converters. The passive method is also applicable for supplying power to small UAVs (Liu and Peng, 2008; Zandi *et al.*, 2011). Directly connected to a DC link, the power sources supply propulsion to the engine based on their own characteristics. By eliminating additional power converters and controllers, the complexity, weight, and power losses can be significantly reduced (Lee *et al.*, 2014).

Using flight testing and power simulations, several researchers have examined and assessed the advantages and disadvantages of active and passive PMSs (Lee *et al.*, 2014; Mudiyanselage *et al.*, 2021; Rajabi *et al.*, 2022). One hybrid power supply system intended for use in low-speed, long-range UAVs was composed of batteries, fuel cells, and solar cells (Lee *et al.,* 2014). During simulation, Lee and coworkers confirmed that a passive power management system (PMS) was unable to maintain the battery's minimum charge level, thus shortening its lifetime and increasing the likelihood of system failure. Conversely, with the use of an active PMS, power losses reached 4.7%. However, in this case the two PMSs were not compared equally: while the passive PMS was only simulated, the active PMS was implemented experimentally. Moreover, many flight-related parameters that could have impacted the data were overlooked.

### a. Rule-based strategies

Based on its simplicity and robustness in management, rule-based control is one of the most prevalent control methods. Due to its comparatively low computing costs, an energy management strategy (EMS) can be implemented online using such a system (Karunarathne *et al.*, 2010a). The PMS that Lee and coworkers used was designed to control a UAV powered by a fuel cell, a battery, and solar cells (Lee *et al.*, 2014). The proposed PMS assigns output power to power sources based on power requirements and battery survivability. The study conducted by Lee et al. relied on solar cells as the principal source of energy because they do not need onboard fuel. DC-DC converters were utilized to set the terminal voltage of each source using the PMS's power outputs as control variables. Fuel cells provided power only during defined power intervals so that the UAV could operate at optimal levels. In tandem with the PMS, a battery management system was used to prevent the battery from overcharging once it reached full charge capacity. For UAVs to operate safely, a battery state of charge of 45% was recommended.

As described in a recent article, Yang et al. devised a state machine strategy for a fuel cell/battery-powered UAV (Yang *et al.*, 2018). Based on battery efficiency and demand power, the logic divides the decision area into five states, which includes two converters, one of which is bidirectional for controlling battery charging and discharging. Additionally, two PI controllers were employed to make voltage and power references. By varying the system's initial conditions, the proposed PMS could be validated by examining the power system's performance. Despite having been implemented in an actual UAV and tested on the ground, this study was not implemented.

Gao and coworkers created an EMS-based long-endurance solar- and battery-powered UAV (Gao *et al.*, 2013). The initial phase involved splitting photovoltaic (PV) energy into three parts: (a) the first part was used to power the UAV, (b) the second part remained in storage for later use, and (c) the third part was employed to charge the battery. Declines in solar irradiance initiated the second phase. Gravitational gliding and stored energy were used to partially address the UAV power deficit. Last, but not least, the battery powered the drone at a low altitude in case of a complete solar power shortage. According to the proposed PMS, the wind effect was also considered in this simulation study, which resulted in about 23.5% greater energy savings compared to another management strategy. Unfortunately, relying on solar power represents a limiting factor for the use of such UAVs.

A rule-based algorithm presented by Savvaris et al. delineated control strategies involving a battery/fuel cell hybrid system (Savvaris *et al.*, 2016). The control variable for this study was power versus current. In addition to relays for power flows, transistors were used for the activation and deactivation of each source. During the power supply operation, three modes were considered: (a) parallel mode, which allows one to use two sources simultaneously; (b) charging mode, which charges the battery; and (c) load the following mode, which considers the load. Additionally, HIL-based experiments were performed. To date, actual flight tests have yet to be conducted with a deployed UAV, which raises the issue of endurance as a possible problem.

In one study, Lee et al. suggested employing constrained thermostat control (CTC) (Lee *et al.*, 2012). As they detailed, solar cells provide the main power source, which can also be used to charge the battery in cases when there might be a surplus of power. Moreover, their developed approach maintains the battery SoC at a low threshold during flight (30%) to facilitate a seamless landing in case the solar cells and the fuel cell are unable to deliver the required power. When the SoC is above the prescribed threshold, the battery contributes to the power supply. To augment the results obtained from this simulation-based study, their findings could be strengthened by conducting a HIL-based approach.

### b. Fuzzy logic strategies

When investigating hybrid power supply systems for UAVs, a fuzzy logic-based PMS should be implemented to enhance power allocation. Fuzzy control algorithms consider inputs such as SoC, power demand (PD), and photovoltaic power, after which they generate control commands accordingly. These algorithms establish priorities and constraints and determine an appropriate management strategy. In general, power generated from either photovoltaic or fuel cells is highly preferred when it comes to supplying power to a UAV. Table 12 provides an example of fuzzification for a battery/fuel cell control system (Zhang, Liu, Dai, *et al.*, 2018). Note that there are three state of charge levels for a battery: low (L), medium (M), and high (H). Additionally, there are five states corresponding to fuzzy power demand (similar to the states for fuzzy output PFC): medium (M), high (H), very high (VH), low (L), and very low (VL) (Erfani and Tavakolan, 2020).

A hybrid fuel cell/battery-propelled UAV powered by an online fuzzy EMS was proposed by Zhang et al. (Zhang, Liu, Dai, *et al.*, 2018). Experiments were conducted to test the designed EMS. A programmable DC/DC converter was used to control the fuel cell's current output, thereby enabling the fuel cell to provide the propulsion. No converter was required since the battery was directly connected to the DC bus. By using the power balance principle in this instance, its output current could be determined indirectly. Utilizing the battery state of charge and power demand as input variables, the implemented algorithm was able to calculate fuel cell power under a fuzzy process. In addition to the fuzzy EMS, the state machine and passive control strategies were compared by assessing three distinct flight missions: pulsed-power missions, flight-power missions, and long-duration missions. Based on fuel consumption and battery state of charge, the proposed strategy was found to perform well. It should be stressed, however, that although the power demand profile was simulated on a test bench using a programmable electric load, no actual flights were conducted to strengthen the findings.

Karunarathne and coworkers developed an energy and power management system with the goal of optimizing the power split between a PEMFC cell and a Li-ion battery (Karunarathne *et al.*, 2010a, 2010b). As they described, EMS is useful for reducing voltage losses caused by oxygen concentration. The control parameters for the PEMFC were determined by the required power, the battery SoC, and the battery control parameters. Next, a short-term implementation can be handled by the PMS once those decisions are made. By enabling the buck/boost mode of the DC/DC bidirectional converter, the PMS controls the PEMFC output through a rule-based system and enables battery charging and discharging. Controlling the inlet air flow rate is achieved using the EMS by regulating the compressor motor voltage on the PEMFC. The membership function estimation parameters can be predicted using an adaptive controller based on ANFIS (Adaptive Neuro-Fuzzy Inference System). To date, this simulation study has yet to be deployed under experimental conditions.

Zhang et al. (2012) and others have proposed combining fuzzy logic and state machines to create a hybrid approach (e.g., Park *et al.*, 2018; Zhang *et al.*, 2018). With respect to UAV batteries and fuel cells, fuzzy logic is preferred; in contrast, state machines are used to manage energy when solar cells and batteries serve as power sources. To examine the behavior of power sources in a simulation platform, a mission scenario was implemented. In comparison with the thermostat control strategy recommended for similar UAVs, the proposed strategy was more effective (Lee *et al.*, 2012). Battery state of charge, fuel consumption, and each source's contribution to the supply process were compared.

Dynamic programming represents a preferred EMS compared to fuzzy logic or rule-based methods (Bradley *et al.*, 2009). Bradley and coworkers tested their proposed algorithm using different flight scenarios, different degrees of hybridization (batteries as a power supply), and with fuel consumption in mind. The various parameters were implemented to determine the optimal architecture in terms of endurance improvements. The main conclusion that emerged from this study is that hybridization may be beneficial in terms of endurance in cases when fuel cells cannot provide the UAV with a reliable source of power.

## Conclusions

Low-altitude air transportation is attracting the attention of a broader swath of logistic companies, online retailers, and governmental agencies. The advantage of drones is that they can avoid geometric obstacles and traffic jams on city streets. Moreover, increasingly urgent environmental factors that include $CO_2$ emission reductions and other energy-saving approaches are intensifying to need to reduce vehicular usage and congestion. When considered in the aggregate, these various factors are advancing the developmental and technological capabilities of UAVs, especially in the roles of monitoring and delivery. For optimal performance and endurance, drones often employ hybrid power supply architecture systems that use some combination of fuel cells, batteries, solar cells, and supercapacitors. To enable today's drones (and those of the future) to work efficiently, the appropriate energy management system must be selected based on optimal and accurate modeling techniques. In addition to assessing the state of the art, this chapter provides insights and recommendations for future research on drone energy supply management and strategy systems.